\documentclass[conference]{IEEEtran}
\IEEEoverridecommandlockouts
\usepackage{cite}
\usepackage{amsmath,amssymb,amsfonts}
\usepackage{algorithmic}
\usepackage{graphicx}
\usepackage{textcomp}
\usepackage{xcolor}
\def\BibTeX{{\rm B\kern-.05em{\sc i\kern-.025em b}\kern-.08em
    T\kern-.1667em\lower.7ex\hbox{E}\kern-.125emX}}
    
\DeclareMathOperator{\Tr}{Tr}
\DeclareMathOperator*{\argmax}{arg\,max}

\usepackage{subcaption}
\usepackage{balance}

\begin{document}

\title{Maximum Likelihood Alternating Summation for Multistatic Angle-based Multitarget Localization}
%MAXIMUM LIKELIHOOD ALTERNATING SUMMATION FOR MULTISTATIC ANGLE-BASED MULTITARGET LOCALIZATION
%Maximum Likelihood Multitarget Localization for Multistatic Angle-based Radars by Alternating Summation
\makeatletter
\newcommand{\newlineauthors}{%
  \end{@IEEEauthorhalign}\hfill\mbox{}\par
  \mbox{}\hfill\begin{@IEEEauthorhalign}
}
\makeatother

\author{\IEEEauthorblockN{Martin Willame}
\IEEEauthorblockA{\textit{ICTEAM-ELEN \& OPERA-WCG}\\
\textit{Université catholique de Louvain} \\
\textit{\& Université libre de Bruxelles} \\
Louvain-la-Neuve \& Brussels, Belgium \\
martin.willame@uclouvain.be}
\and
\IEEEauthorblockN{Laurent Storrer}
\IEEEauthorblockA{\textit{OPERA-WCG} \\
\textit{Université libre de Bruxelles}\\
Brussels, Belgium \\
laurent.storrer@ulb.be}
\and
\IEEEauthorblockN{Hasan Can Yildirim}
\IEEEauthorblockA{\textit{OPERA-WCG}\\
\textit{Université libre de Bruxelles} \\
Brussels, Belgium \\
hasan.can.yildirim@ulb.be}
\newlineauthors
\IEEEauthorblockN{François Horlin}
\IEEEauthorblockA{\textit{OPERA-WCG} \\
\textit{Université libre de Bruxelles}\\
Brussels, Belgium \\
francois.horlin@ulb.be}
\and
\IEEEauthorblockN{Jérôme Louveaux}
\IEEEauthorblockA{\textit{ICTEAM-ELEN} \\
\textit{Université catholique de Louvain}\\
Louvain-la-Neuve, Belgium \\
jerome.louveaux@uclouvain.be}
}

\maketitle

\begin{abstract}
Recent advancements in Wi-Fi sensing have sparked interest in exploiting OFDM modulated communication signals for target detection and tracking. In this study, we address the angle-based localization of multiple targets using a multistatic OFDM radar. While the maximum likelihood approach optimally merges data from each radar pair comprised by the system, it entails a complex multi-dimensional search process. Leveraging pre-estimation of the targets' parameters obtained via the MUSIC algorithm, our method decouples this multi-dimensional search into a single two-dimensional estimator per target. The proposed alternating summation method allows the computation of a combined likelihood map aggregating contributions from each radar pair, enabling target detection via peak selection. Besides reducing computational complexity, the method effectively captures target interactions and accommodates varying radar pair localization abilities. Also, it requires transmitting only the estimated channel covariance matrices of each radar pair to the central processor. Numerical simulations demonstrate superior performance over existing approaches.

\end{abstract}

\vskip0.5\baselineskip

\section{Introduction} \label{sec:intro}
In recent years, Wi-Fi sensing has gained significant attention in the literature. The known preambles of orthogonal frequency division multiplexing (OFDM) modulated communication signals are exploited for target detection and tracking \cite{9477585}. Within this context, the Wi-Fi transmitter is denoted as the sensing transmitter (STx), while signal processing is undertaken by the sensing receiver (SRx). When these components are respectively collocated or not, they constitute a monostatic or bistatic radar pair (RP). A multistatic radar system is characterized by the integration of multiple RPs observing the same scene. Through the aggregation of information from each RP by a central processor, the system sensing capacity is enhanced \cite{554205}. This fusion can occur at various levels.  Parametric level integration involves the aggregation of local decisions regarding the target position, while signal level fusion entails the combination of raw signals from each RP to estimate the target position. Although signal level fusion enhances sensing capabilities by exploiting all available information, it often necessitates robust backhaul links and entails increased computational costs.

In this study, we examine the localization of $K$ targets using a multistatic OFDM radar system that analyzes the angle-of-departure (AoD) and angle-of-arrival (AoA) at each RP. We focus on the development of signal level fusion algorithms with limited data transfer and computational complexity to circumvent these main drawbacks.

The maximum likelihood (ML) approach enables the derivation of an optimal signal level fusion rule. However, in scenarios involving the detection of multiple targets, the received signals reflected by the $K$ targets are not orthogonal. Consequently, the complexity of the ML algorithm increases, as its brute force implementation necessitates a $2K$-dimensional search across the positions of all targets. In \cite{7543}, the authors introduce the alternating projection method, which efficiently computes the ML estimator for a single RP. From an initial estimate of the $K$ target positions, the method iteratively maximizes the likelihood with respect to a single parameter while holding others fixed. This approach adeptly addresses the non-orthogonality of the signals reflected by the targets. Nonetheless, extending this method to a multistatic radar configuration poses significant challenges. To the best of the author's knowledge, resolving this issue remains an open problem.

Besides the ML, angle processing can also be performed through subspace-based methods. For instance, the multiple signal classification (MUSIC) algorithm enables a two-dimensional search by separating the signal from the noise subspace \cite{1143830}. In our prior investigations \cite{willame2024multistatic}, we leveraged the connection between MUSIC and ML estimators to approximate the multidimensional ML parameter estimation through a weighted combination of MUSIC pseudo-spectrum from multiple RPs. This reduces the complexity of the $2K$-dimensional search of the ML estimator into $K$ two-dimensional problems solved by MUSIC. The proposed fusion method for joint AoD/AoA-based localization of $K$ targets demonstrates superior performance compared to other parametric level fusion methods relying on MUSIC outputs. Despite its promising outcomes, the method is suboptimal due to its uniform weighting of the MUSIC pseudo-spectrum outputs. Consequently, it fails to account for the RP's varying ability to localize different targets owing to geometric considerations or disparate path losses. 

In this work, we propose to address this issue by developing a per-target ML fusion of the different RPs via alternating summation. The interaction among the different targets is captured by relying on a pre-estimation of the targets' AoD/AoA obtained through the MUSIC algorithm. These per-target likelihood functions are then aggregated into a single detection map, upon which a peak detector is applied.

The paper is structured as follows. Section \ref{sec:system_model} presents the system model. Section \ref{sec:MLF} details the mathematical formulation and discussion of the proposed ML combination rule. Section \ref{sec:sim_res} provides numerical results to evaluate the effectiveness of the novel fusion rule.
\subsection{Major Contributions}
Our contributions are outlined as follows:
\begin{itemize}
\item We introduce a signal-level fusion methodology based on the ML framework for joint AoD/AoA-based localization of $K$ targets by a multistatic OFDM radar. Based on pre-estimation of the targets' parameters acquired via MUSIC, the proposed alternating summation method effectively decouples the $2K$-dimensional ML estimator into $K$ per-target two-dimensional searches.
\item Differing from our prior work \cite{willame2024multistatic}, this methodology adeptly captures the interaction among the targets and account for the varying localization ability of the RPs, all without increasing the dimensionality of the search.
\item Numerical simulations demonstrate the advantages of this method compared to other approaches, and investigate the influence of the signal-to-noise ratio (SNR) on the localization performance of a multistatic radar composed of two OFDM radar pairs.
\end{itemize}

\subsection{Notations}
The vectors and matrices are defined as $\mathbf{a}$ and $\mathbf{A}$, respectively. The trace, the transpose and the Hermitian transpose are denoted $\Tr\left\{\mathbf{A}\right\} ,\mathbf{A}^{\mathrm T}$ and $\mathbf{A}^\textup{H}$, respectively. The Moore-Penrose inverse is defined as $\mathbf{A}^{+} =\left(\mathbf{A}^\textup{H} \mathbf{A} \right)^{-1} \mathbf{A}^\textup{H}$. The Kronecker product is denoted $\otimes$.

%Besides the ML, angle processing can also be performed based on subspace-based methods. For instance, the multiple signal classification (MUSIC) algorithm allows a two-dimensional search by separating the signal from the noise subspace \cite{1143830}. In our prior investigations \cite{willame2024multistatic}, we leveraged the connection between MUSIC and ML estimators to approximate the multidimensional ML parameter estimation through a weighted combination of MUSIC pseudo-spectrum from multiple RPs. The proposed fusion method for joint AoD/AoA-based localization of $K$ targets demonstrates superior performance compared to other parametric level fusion methods relying on MUSIC outputs. It offers three main advantages: (i) applicability to OFDM multistatic radar systems with direct possible extension to other radar configurations, (ii) reduction of the ML estimator's complexity from a $2K$-dimensional search to $K$ two-dimensional problems, and (iii) processing solely relies on the transmission of the sample covariance matrix of the estimated channel by each RP to the central processor. Despite its promising outcomes, the method is suboptimal due to its uniform weighting of the MUSIC pseudo-spectrum outputs. Consequently, it fails to account for the RP's varying ability to localize different targets owing to geometric considerations or disparate path losses.
\section{System Model} \label{sec:system_model}
In this study, we investigate a multistatic OFDM radar setup designed for target localization in the $x$-$y$ plane. The configuration consists of multiple STx-SRx RPs, where the data processed by each RP is transmitted to a central processor. Subsequently, this data is combined to ascertain the target positions. Details regarding the data transmission and the fusion rule employed at the central processor are elaborated in Section \ref{sec:MLF}.

To express the channel model, we make the following assumptions.
\begin{enumerate}
\item The OFDM symbols transmitted by distinct radar pairs maintain orthogonality, achieved through frequency or time division multiple access techniques. This orthogonality permits each SRx to process the frame transmitted by its corresponding STx devoid of interference from other STx.
\item The direct signal between the STx and the SRx, along with clutter contributions, are effectively suppressed from the estimated channel.
\item Only multipath signals featuring a single reflection on a target are significantly impacting the observed channel model. Signals with multiple reflections are are thus omitted.
\end{enumerate}

\begin{figure}
\centering
\includegraphics[width=0.95\columnwidth]{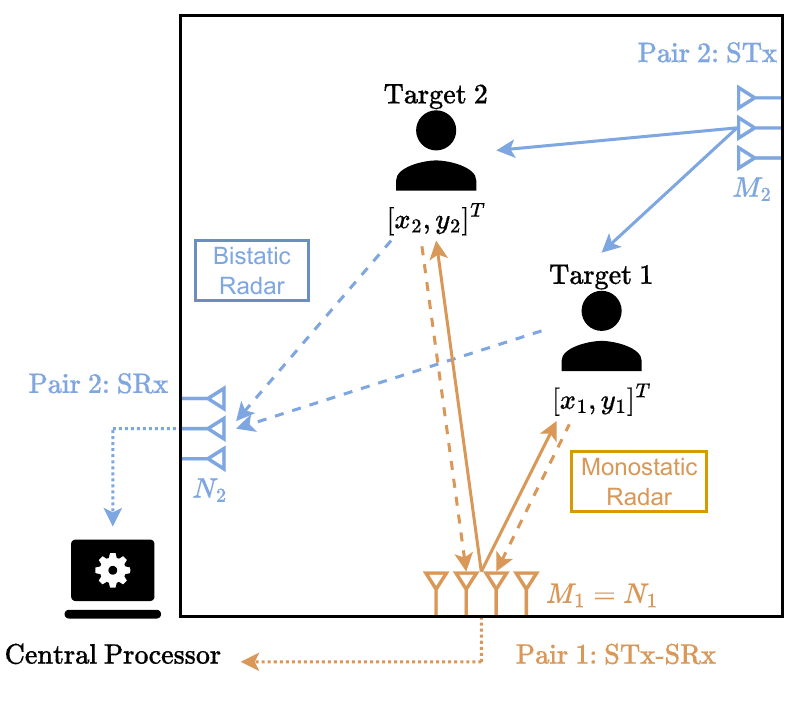}
\caption{Illustration of the scenario. The multistatic configuration comprises a bistatic radar and a monostatic radar. In the illustration, solid lines represent the incident waveforms, dashed lines represent the reflected waveforms, and dotted lines depict the data transmitted to the central processor.}
\label{fig:scenario}
\end{figure}

The scenario is depicted in \figurename~\ref{fig:scenario}, where the multistatic radar system aims at localizing $K$ targets within its coverage area. The positions of these $K$ targets are defined by the vectors $\mathbf{x}=[x_1 \ \dots \ x_K]^{\mathrm T}$ and $\mathbf{y}=[y_1 \ \dots \ y_K]^{\mathrm T}$. The system comprises $P$ RPs, each consisting of an STx and an SRx equipped with a uniform linear array. For simplicity, we assume the array has half-wavelength spacing and is oriented towards the coverage area. In the $p\textsuperscript{th}$ radar pair, the AoD and AoA for the $k\textsuperscript{th}$ target are denoted by $\varphi_{p,k}$ and $\vartheta_{p,k}$, respectively. These angles are defined between the wavefront and the normal vector of the corresponding antenna array. The sets of all AoDs and AoAs are denoted by $\boldsymbol \Phi_p$ and $\boldsymbol \Theta_p$, respectively. Throughout this paper, for notation simplicity, we define the joint AoD/AoA set as $\boldsymbol \Psi_p = (\boldsymbol\Phi_p,\boldsymbol\Theta_p)$ and do not explicitly write its dependence on $(\mathbf{x},\mathbf{y})$. The $(M_p \times 1)$ AoD and $(N_p \times 1)$ AoA steering vectors are thus given by
\begin{align}
\mathbf{v}_p(\varphi_{p,k})&= [1 \ e^{j\pi \sin(\varphi_{p,k})} \dots \ e^{j\pi (M_p-1) \sin(\varphi_{p,k})}]^{\mathrm T},\\
\mathbf{v}_p(\vartheta_{p,k})&= [1 \ e^{j\pi \sin(\vartheta_{p,k})} \dots \ e^{j\pi (N_p-1) \sin(\vartheta_{p,k})}]^{\mathrm T},
\end{align}
where $M_p$ and $N_p$ denote the number of transmitting and receiving antennas, respectively. The STx of the $p\textsuperscript{th}$ RP transmits $M_p$ OFDM symbols isotropically with $Q_p$ subcarriers and a subcarrier spacing of $\Delta_f$. 

The baseband equivalent channel matrix for the $p\textsuperscript{th}$ RP is stacked into a vector for each subcarrier $q$ as follows:
\begin{equation} \label{eq:channel}
\mathbf{h}_{p,q}(\boldsymbol\Psi_p) = \mathbf{A}_p(\boldsymbol\Psi_p) \ \boldsymbol\alpha_{p,q},
\end{equation}
where
\begin{itemize}
    \item $\mathbf{h}_{p,q}$ is the $M_p N_p \times 1$ channel vector of the $p\textsuperscript{th}$ RP for subcarrier $q$.
    \item $\mathbf{A}_p(\boldsymbol\Psi_p) = \mathbf{A}_p(\boldsymbol\Phi_p,\boldsymbol\Theta_p) = [\mathbf{a}_{p,1}(\varphi,\vartheta) \ \dots \ \mathbf{a}_{p,K}(\varphi,\vartheta)]$ represents the joint AoD/AoA steering matrix  ($M_p N_p \times K$). Here, the short notation $\mathbf{a}_{p,k}(\varphi,\vartheta)$ denotes the joint AoD/AoA steering vector defined as $\mathbf{a}_p(\varphi_{p,k},\vartheta_{p,k}) = \mathbf{v}_p(\varphi_{p,k}) \otimes \mathbf{v}_p(\vartheta_{p,k})$.
    \item $\boldsymbol\alpha_{p,q}=[\alpha_{p,q,1} \ \dots \ \alpha_{p,q,K}]^{\mathrm T}$ is the channel coefficient vector ($K \times 1$). These coefficients include the linearly increasing phases across subcarriers due to the range of the targets and the attenuation defined by the radar range equation \cite{101049}. Any stochastic model can be associated with the channel coefficient (e.g. the Swerling model \cite{1057561}).
\end{itemize}
\section{Maximum Likelihood Fusion}\label{sec:MLF}
In this section, we formulate the ML combination rule for determining the positions of the $K$ targets based on the signal acquired by each RP. As discussed in Section \ref{sec:intro}, we develop a per-target likelihood function at each RP and define a combination rule to account for the varying localization ability of the RP without relying on the $2K$-dimensional ML search.

The parameters to be estimated are defined by the vector $\boldsymbol \gamma = [\mathbf{x}^{\mathrm{T}} \ \mathbf{y}^{\mathrm{T}} \ \{ \boldsymbol\alpha_{p,0}^{\mathrm{T}} \ \dots \ \boldsymbol\alpha_{p,Q_p-1}^{\mathrm{T}} \}_{p=1\dots P}]^{\mathrm{T}}$. It is assumed that the number of targets to be localized is known as methods for estimating $K$ are available in the literature \cite{1164557}. At each RP, information about the targets is acquired by estimating the channel at the SRx using $M_p$ known OFDM symbols transmitted by the STx. Consequently, the observed data from the $p\textsuperscript{th}$ RP for ML development is modeled as a noisy channel estimated vector
\begin{equation} \label{eq:observations}
\widetilde{\mathbf{h}}_{p,q} = \mathbf{A}_p(\boldsymbol\Psi_p) \ \boldsymbol\alpha_{p,q} + \mathbf{n}_{p,q}, 
\end{equation}
where the $(M_p N_p \times 1)$ vector $\mathbf{n}_{p,q}$ denotes the estimation errors, which are assumed to be Additive White Gaussian Noise (AWGN) with variance $\sigma^2_p$.

Notice that our study solely focuses on joint AoD/AoA-based ML localization and does not utilize the range information. As a result, each channel coefficient vector $\boldsymbol\alpha_{p,q}$ is independently estimated, since the linear phase increase across subcarriers, defined by the range of each target, is not exploited.

Considering independent noise contributions for the estimated channel vector from all radar pairs, the combined likelihood function is derived as the product of individual Gaussian density functions. After taking the natural logarithm of the combined likelihood function, the sum of the local log-likelihood functions, which are denoted as $\mathcal{L}_p(\boldsymbol\gamma)$, has to be maximized,
\begin{equation} \label{eq:global_max}
\hat{\boldsymbol \gamma} = \argmax_{\boldsymbol \gamma} \mathcal{L}(\boldsymbol \gamma) = \argmax_{\boldsymbol \gamma} \sum_{p=1}^P \mathcal{L}_p(\boldsymbol \gamma).
\end{equation}
These local log-likelihood functions are developed in the subsequent subsection separately for each radar pair, as their contributions are independent.

\subsection{Development of the Local Likelihood Functions} \label{sec:local_ML_function}
The local log-likelihood function for the $p\textsuperscript{th}$ radar pair is given by
\begin{equation} \label{eq:ind_log_likelihood}
\mathcal{L}_p(\boldsymbol \gamma) =  \frac{-1}{2\sigma_p^2} \sum_{q=0}^{Q_p-1} \lVert \widetilde{\mathbf{h}}_{p,q} - \mathbf{A}_p(\boldsymbol\Psi_p) \ \boldsymbol\alpha_{p,q} \rVert^2.
\end{equation}
First, we maximize with respect to the channel coefficients $\boldsymbol\alpha_{p,q}$ to obtain a closed-form expression as a function of $\boldsymbol\Psi_p$. Following the solution of the resulting linear least squares problem, the ML estimate of the channel coefficients for every subcarrier $q$ is expressed by
\begin{equation} \label{eq:Moore-Penrose}
\Hat{\boldsymbol\alpha}_{p,q}(\boldsymbol\Psi_p) = \mathbf{A}_p^{+}(\boldsymbol\Psi_p)\ \widetilde{\mathbf{h}}_{p,q}.
\end{equation}
Substituting these estimates \eqref{eq:Moore-Penrose} back into \eqref{eq:ind_log_likelihood} and after performing some mathematical manipulations, the local log-likelihood function can be reformulated as 
\begin{equation} \label{eq:trace_log_likelihood}
\mathcal{L}_p(\boldsymbol \gamma) = \frac{Q_p}{2\sigma_p^2} \Tr\left\{ \mathbf{A}_p(\boldsymbol\Psi_p)\mathbf{A}_p^{+}(\boldsymbol\Psi_p) \ \widetilde{\mathbf{R}}_p\right\},
\end{equation}
in which $\widetilde{\mathbf{R}}_p = \frac{1}{Q_p} \sum_{q=0}^{Q_p-1} \widetilde{\mathbf{h}}_{p,q} ~\widetilde{\mathbf{h}}^\textup{H}_{p,q}$ denotes the sample covariance matrix of the channel vector averaged over the subcarriers. We seek the set of $(x,y)$ positions of the $K$ targets that maximizes the sum of the local log-likelihood functions in \eqref{eq:global_max}. Therefore, the brute-force maximization of the sum of \eqref{eq:trace_log_likelihood} implies solving a $2K$-dimensional problem due to the presence of multiple targets within the coverage area.

\subsection{Decoupling the Maximum Likelihood}
\label{sec:ML_Layers}
In this subsection, we demonstrate that if each RP can pre-estimate the target parameters $\Hat{\mathbf{\Psi}}_p=(\Hat{\boldsymbol \Phi}_p,\Hat{\boldsymbol \Theta}_p)$, then the $2K$-dimensional local log-likelihood functions can be approximately decoupled into $K$ per-target two-dimensional functions. 

This pre-estimation can be independently obtained at each RP from their MUSIC spectrum, as given by
\begin{equation} \label{eq:MUSIC spectrum}
\mathcal{J}_p(\varphi,\theta) = \frac{1}{\mathbf{a}_p^\textup{H}(\varphi,\theta)~ \widetilde{\mathbf{G}}_p \widetilde{\mathbf{G}}_p^\textup{H} ~\mathbf{a}_p(\varphi,\theta)},
\end{equation}
where $\widetilde{\mathbf{G}}_p$  represents the estimated noise subspace matrix obtained from the singular value decomposition of $\widetilde{\mathbf{R}}_p$. The $K$ eigenvectors corresponding to the strongest eigenvalues form the signal subspace, while the remaining vectors form the noise subspace \cite{1143830}. The $K$ highest peaks in $\mathcal{J}_p(\varphi,\theta)$ are denoted as $\Hat{\varphi}_{p,k},\Hat{\theta}_{p,k}$ corresponding to each target $k$, thereby constituting the pre-estimated set of joint AoD/AoA $\Hat{\mathbf{\Psi}}_p=\{(\Hat{\varphi}_{p,k},\Hat{\theta}_{p,k})_{k=1\dots K}\}$.

Once each RP possesses a reliable initial estimation $\Hat{\mathbf{\Psi}}_p$ of the joint AoD/AoA of the $K$ targets, we introduce $\Hat{\mathbf{\Psi}}_p^{(k)}(x_k,y_k)$
where $\Hat{\varphi}_{p,k}$ and $\Hat{\theta}_{p,k}$ are replaced by the unknown angles $\varphi_{p,k}(x_k,y_k)$ and $\theta_{p,k}(x_k,y_k)$, while the $K-1$ other angles are held fixed to their pre-estimated values. For notation simplicity, we will omit the dependence of $\Hat{\mathbf{\Psi}}_p^{(k)}$ on $(x_k,y_k)$. Leveraging \eqref{eq:trace_log_likelihood} and this new definition, the local per-target log-likelihood function for $k = 1 \dots K$ is defined as
\begin{equation} \label{eq:local_individual_log_individual}
\mathcal{L}_{p}^{(k)}(x_k,y_k) = \frac{Q_p}{2\sigma_p^2} \Tr\left\{ \mathbf{A}_p(\Hat{\mathbf{\Psi}}_p^{(k)})\mathbf{A}_p^{+}(\Hat{\mathbf{\Psi}}_p^{(k)}) \ \widetilde{\mathbf{R}}_p\right\}.
\end{equation}
Due to the pre-estimation of the joint AoD/AoA of the other $K-1$ targets, $\mathcal{L}_{p}^{(k)}$ solely depends on the position $(x_k,y_k)$ of the $k\textsuperscript{th}$ target. This enables the decoupling of the $2K$-dimensional local log-likelihood function into $K$ per-target two-dimensional functions. 

From \eqref{eq:local_individual_log_individual}, a local $x$-$y$ detection map per RP, representing the log-likelihood of any of the $K$ targets being present at a given position $(x,y)$, can be established. Since it is unknown which target is being detected at a particular position, the most probable target is the one with the highest per-target log-likelihood. Consequently, the local log-likelihood function $\mathcal{L}_{p}(x,y)$ corresponding to the presence of any of the $K$ targets at a given $(x,y)$ position is described by
\begin{align} 
\mathcal{L}_{p}(x,y) 
& = \max_{k} \mathcal{L}_{p}^{(k)}(x,y) \\
& = \max_{k} \frac{Q_p}{2\sigma_p^2} \Tr\left\{ \mathbf{A}_p(\Hat{\mathbf{\Psi}}_p^{(k)})\mathbf{A}_p^{+}(\Hat{\mathbf{\Psi}}_p^{(k)}) \ \widetilde{\mathbf{R}}_p\right\} \label{eq:local_log}
\end{align}

\subsection{Fusion of Multiple Radar Pairs} \label{sec:fusion}
In order to leverage the spatial diversity inherent in the multistatic system, the targets' positions are inferred from the combined log-likelihood functions of all RPs defined in \eqref{eq:global_max}. Formulating a per-target combined log-likelihood function requires an association step between the targets' parameters pre-estimated at each RP. This introduces complexity, particularly in scenarios where not all RPs receive significant contributions from the $K$ targets or when certain pre-estimations are misleading.

To circumvent this association step, we propose determining the targets' positions by identifying the $K$ highest peaks in a combined $x$-$y$ detection map. Based on the local log-likelihood function for each RP provided in \eqref{eq:local_log}, the combined log-likelihood function for a given position $(x,y)$ in the detection map is given by
\begin{equation} \label{eq:combined_log}
\mathcal{L}(x,y) 
= \sum_{p=1}^P \mathcal{L}_{p}(x,y).
\end{equation} 
The proposed method is described as an alternating summation, as the per-target likelihood function maximizing \eqref{eq:local_log} at each RP may vary for different points $(x,y)$.

The pre-estimation by the MUSIC algorithm enabling the evaluation of \eqref{eq:combined_log}, followed by a highest peak selection, effectively replaces the $2K$-dimensional ML parameter estimation. The evaluation of the combined log-likelihood can be performed by the central processor solely using the sample covariance matrix $\Tilde{\mathbf{R}}_p$ from every RP. Hence, the proposed signal fusion algorithm requires each RP $p$ to transmit an $(M_p N_p)$ square matrix to the central processor, instead of a $(Q_p \times M_p \times N_p)$ estimated channel tensor.

\section{Simulation Results} \label{sec:sim_res}
\begin{figure}
\centering
\includegraphics[trim=3.4cm 9.5cm 4.3cm 9.91cm,width=0.95\columnwidth]{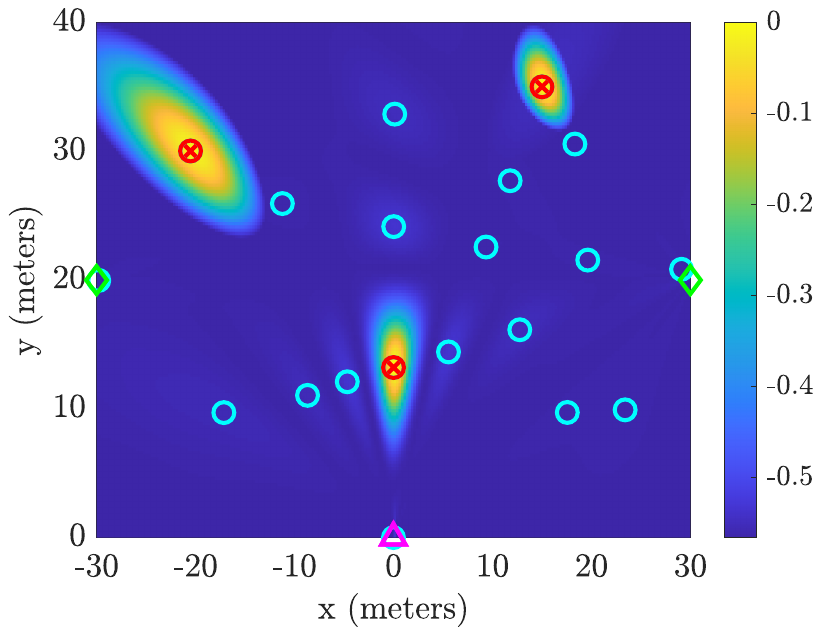}
\caption{
The combined log-likelihood map is depicted in dB and normalized to the value of the highest peak. The two STx are symbolized by diamonds, the SRx by a triangle, the true target positions by crosses, the estimated locations of the targets by red circles, and the other peaks in the map by blue circles.}
\label{fig:setup}
\end{figure}

\begin{figure*}[!t]
\centering
\subfloat[RMSE]{\includegraphics[trim=3.4cm 9.5cm 4.3cm 9.91cm,width = 0.42\textwidth]{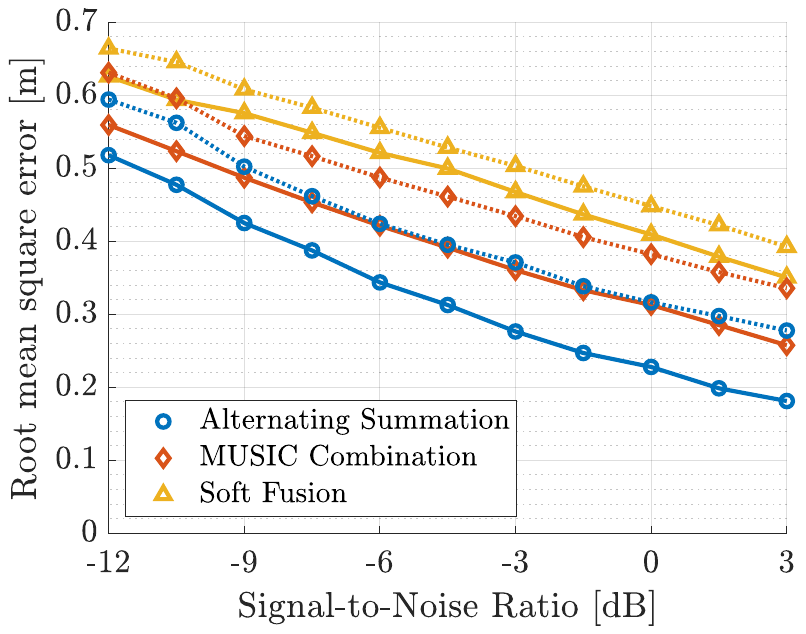}}
\label{fig:graphe1a}
\hfill
\subfloat[Hit Rate]{\includegraphics[trim=3.4cm 9.5cm 4.3cm 9.91cm,width = 0.42\textwidth]{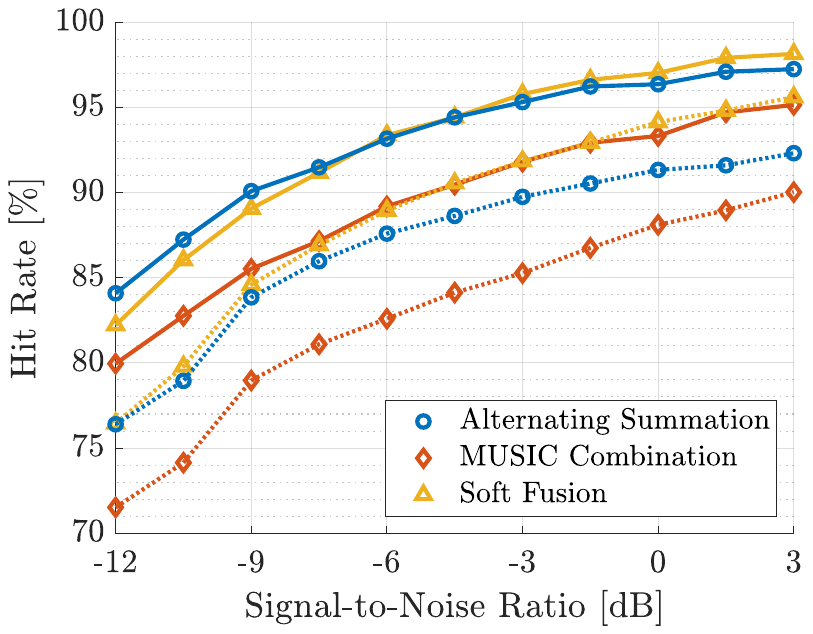}}
\label{fig:graphe1b}
\caption{The impact of the signal-to-noise ratio on the performance metrics of the various methods is depicted. The solid lines correspond to a scenario with three targets, while the dashed lines represent a scenario with six targets.}
\label{fig:graphe1}
\end{figure*}

This section evaluates the enhancement brought by the proposed alternating summation fusion method on the accuracy of a multistatic radar system in localizing targets within the coverage area. \figurename~\ref{fig:setup} illustrates the setup for a single realization of a scenario with three targets and displays the combined log-likelihood map obtained using the method proposed in Section \ref{sec:MLF}. It is noteworthy that what is typically interpreted as a noise floor in classical detection maps results from the combined log-likelihood computed with $K-1$ pre-estimated targets. The height of the peaks indicates the extent to which a target at a given position enhances this likelihood.

The method outlined in Section \ref{sec:MLF} is compared through $15,000$ Monte-Carlo simulations with a soft fusion and the MUSIC combination presented in our prior work \cite{willame2024multistatic}.  The following performance metrics are studied:
\begin{enumerate}
\item \textbf{Hit Rate:} A hit is acknowledged when a peak is detected within a vicinity of 2 meters from the true target position; otherwise, it is considered a miss.
\item \textbf{RMSE:} The root mean square error (RMSE) is computed between the true target position and the corresponding detected position. To ensure a fair comparison between the methods, only the targets resulting in a hit by all methods are considered in the evaluation of the RMSE.
\end{enumerate}

\balance For each simulation, we analyze a fixed multistatic radar setup comprising two STx and one SRx, resulting in two RPs designated to locate randomly positioned targets.  Initially, peak detection is executed on an $x$-$y$ grid with a spacing of $0.25$ meters, followed by refinement using gradient descent for each detected peak to enable off-grid detections. Similarly, a gradient descent method is employed for peak detection in the local MUSIC spectrums. 

\figurename~\ref{fig:graphe1} depicts the results obtained for the hit rate and the RMSE as functions of the signal-to-noise ratio (SNR) in scenarios with three and six targets, represented by the solid and dashed lines respectively. The SNR is defined here as the mean of the quotient of the channel coefficients of each path by the noise variance. It is observed that in both scenarios, the proposed method enhances the achievable RMSE compared to existing detection algorithms. Similarly, the hit rate is also improved compared to the other methods. However, the soft fusion exhibits better hit rates as the number of targets and the SNR increases. This is explained by the fact that in the simulations, the association step between the detections of each RP, required by the soft fusion method, is assumed to be perfect. In practice, the inability of the soft fusion to distinguish unmatched local detections would degrade the localization performance. 
\section{Conclusion}
In this paper, we introduce a novel data-level fusion method derived from the ML framework for a multistatic OFDM radar system aimed at jointly localizing $K$ targets based on their joint AoD/AoA. The proposed alternating summation method exploits the MUSIC algorithm for pre-estimating the targets' parameters, enabling the decoupling of the $2K$-dimensional search required by the ML estimator into $K$ per-target two-dimensional likelihood functions. The estimation of the targets' positions is then achieved through a highest peak selection on a combined likelihood map. This process can be executed by the central processor using solely the sample covariance matrix from each radar pair, thus eliminating the need for transmitting the full raw signal observations. Monte-Carlo simulations demonstrate that the proposed method outperforms existing localization algorithms in the literature by effectively addressing the varying localization capabilities of the radar pairs.
In future works, the proposed methodology could be further improved by incorporating accurate data association techniques for the pre-estimations obtained from different radar pairs.

\bibliographystyle{IEEEtran}
\bibliography{IEEEabrv,References}

\begin{thebibliography}{1}
\providecommand{\url}[1]{#1}
\csname url@samestyle\endcsname
\providecommand{\newblock}{\relax}
\providecommand{\bibinfo}[2]{#2}
\providecommand{\BIBentrySTDinterwordspacing}{\spaceskip=0pt\relax}
\providecommand{\BIBentryALTinterwordstretchfactor}{4}
\providecommand{\BIBentryALTinterwordspacing}{\spaceskip=\fontdimen2\font plus
\BIBentryALTinterwordstretchfactor\fontdimen3\font minus
  \fontdimen4\font\relax}
\providecommand{\BIBforeignlanguage}[2]{{%
\expandafter\ifx\csname l@#1\endcsname\relax
\typeout{** WARNING: IEEEtran.bst: No hyphenation pattern has been}%
\typeout{** loaded for the language `#1'. Using the pattern for}%
\typeout{** the default language instead.}%
\else
\language=\csname l@#1\endcsname
\fi
#2}}
\providecommand{\BIBdecl}{\relax}
\BIBdecl

\bibitem{9477585}
L.~Storrer, H.~C. Yildirim, M.~Crauwels, E.~I.~P. Copa, S.~Pollin, J.~Louveaux,
  P.~De~Doncker, and F.~Horlin, ``Indoor tracking of multiple individuals with
  an 802.11ax wi-fi-based multi-antenna passive radar,'' \emph{IEEE Sensors
  Journal}, vol.~21, no.~18, pp. 20\,462--20\,474, 2021.

\bibitem{554205}
D.~Hall and J.~Llinas, ``An introduction to multisensor data fusion,''
  \emph{Proceedings of the IEEE}, vol.~85, no.~1, pp. 6--23, 1997.

\bibitem{7543}
I.~Ziskind and M.~Wax, ``Maximum likelihood localization of multiple sources by
  alternating projection,'' \emph{IEEE Transactions on Acoustics, Speech, and
  Signal Processing}, vol.~36, no.~10, pp. 1553--1560, 1988.

\bibitem{1143830}
R.~Schmidt, ``Multiple emitter location and signal parameter estimation,''
  \emph{IEEE Transactions on Antennas and Propagation}, vol.~34, no.~3, pp.
  276--280, 1986.

\bibitem{willame2024multistatic}
M.~Willame, H.~Yildirim, L.~Storrer, F.~Horlin, and J.~Louveaux, ``Multistatic
  ofdm radar fusion of music-based angle estimation,'' 2024.

\bibitem{101049}
M.~Richards, J.~Scheer, and W.~Holm, \emph{Principles of Modern Radar}.\hskip
  1em plus 0.5em minus 0.4em\relax SciTech Pub., 2010, no. vol.~3.

\bibitem{1057561}
P.~Swerling, ``Probability of detection for fluctuating targets,'' \emph{IRE
  Transactions on Information Theory}, vol.~6, no.~2, pp. 269--308, 1960.

\bibitem{1164557}
M.~Wax and T.~Kailath, ``Detection of signals by information theoretic
  criteria,'' \emph{IEEE Transactions on Acoustics, Speech, and Signal
  Processing}, vol.~33, no.~2, pp. 387--392, 1985.

\end{thebibliography}
\end{document}